# Image-Guided Microstructure Optimization using Diffusion Models: Validated with Li-Mn-rich Cathode Precursors

Geunho Choi,[‡] Changhwan Lee,[‡] Jieun Kim, Insoo Ye, Keeyoung Jung[*] and Inchul Park[*]

**Abstract**

Microstructure often dictates materials performance, yet it is rarely treated as an explicit design variable because microstructure is hard to quantify, predict, and optimize. Here, we introduce an image centric, closed-loop framework that makes microstructural morphology into a controllable objective and demonstrate its use case with Li- and Mn-rich layered oxide cathode precursors. This work presents an integrated, AI driven framework for the predictive design and optimization of lithium-ion battery cathode precursor synthesis. This framework integrates a diffusion-based image generation model, a quantitative image analysis pipeline, and a particle swarm optimization (PSO) algorithm. By extracting key morphological descriptors such as texture, sphericity, and median particle size ($D_{50}$) from SEM images, the platform accurately predicts SEM like morphologies resulting from specific coprecipitation conditions, including reaction time-, solution concentration-, and pH-dependent structural changes. Optimization then pinpoints synthesis parameters that yield user defined target morphologies, as experimentally validated by the close agreement between predicted and synthesized structures. This framework offers a practical strategy for data driven materials design, enabling both forward prediction and inverse design of synthesis conditions and paving the way toward autonomous, image guided microstructure engineering.

## Introduction

Microstructural morphology plays a pivotal role in material performance by orchestrating complex phases and controlling the heterogeneity of microstructural features.[1] Transformation-Induced Plasticity (TRIP) steels,[2] for instance, feature a meticulously crafted microstructure that undergoes deformation-driven phase transitions, yielding multiple coexisting phases. The resulting mechanical properties are a product of synergistic interactions, resulting in strengths and ductilities that far surpass the mere sum of each phase's contribution. In case of the cathode active materials for Li-ion batteries, individual particles are never identical; thus, morphological heterogeneity likewise dominates performance. For Li- and Mn--rich (LMR) layered oxide cathodes, tailored composition can provide an initial capacity of over 250 mAh g$^{-1}$; however, long-term stability and voltage retention are primarily influenced by variations in particle size, morphology, and defect concentration.[3] The inherent complexity of LMR microstructures has, in fact, obscured the critical link between morphology, structural defects, and reaction mechanisms, thereby highlighting a persistent challenge in materials design.[4-6] While composition and scalar parameters are routinely optimized, the rich morphology captured in imaging data remains an underexploited design lever.

Conventional materials optimization frameworks generally address microstructure in a relatively indirect manner, with a focus on maximizing scalar targets (e.g., capacity, conductivity) or adjusting composition and stoichiometry.[5] This approach is often predicated on the assumption that microstructural details will emerge because of processing-related decisions. Consequently, electron microscopic image-level features are seldom explicitly delineated as design objectives. In the domain of battery cathode development, for instance, synthesis recipes are optimized to enhance electrochemical properties. The microstructure is evaluated post hoc using bulk average properties such as XRD phase refinement, BET surface area, particle size distribution, and porosity.[4] This approach contrasts with the more conventional practice of designing these proxies upfront. The quantification of an entire SEM or TEM image in a form suitable for optimization remains a non-trivial task. Previous studies have employed a limited set of summary metrics, such as phase area fraction or characteristic length scales to incorporate morphology into design.[7, 8] However, these coarse descriptors may overlook the subtle textural and topological nuances that are crucial for performance. Furthermore, there is a lack of discussion regarding the sufficiency of indirect measures, resulting in a suboptimal representation of rich, image-derived information within the inverse design loop.

In addition to the limitations of coarse descriptors, effective image-based inverse design of microstructures confronts several intertwined challenges. The process of distilling the rich, multi-scale patterns visible in SEM and TEM images into quantitative features is exceptionally difficult. Conventional metrics such as median grain size, porosity, or aspect ratio capture only fragmentary aspects of morphology, leaving subtle textural and topological nuances unquantified. The establishment of robust, predictive links between synthesis parameters and the resulting microstructure remains largely empirical. Minor changes in precursor concentration, temperature, or mixing protocol can provoke disproportionate or unpredictable morphological shifts. This forces reliance on laborious trial-and-error tuning. Absent an integrated closed-loop framework that treats image-derived morphology as an explicit design variable, researchers resort to an inefficient Edisonian cycle adjust, synthesize, characterize, and repeat an approach that becomes untenable as materials systems become more complex. In order to surmount the aforementioned impediments, it is imperative to employ methodologies that systematically encode substantial image data, facilitate dependable process-morphology predictions, and underpin autonomous, closed-loop design of targeted microstructures.

To address these gaps, we propose a closed-loop, image-driven inverse design framework that integrates wavelet-based image quantification, diffusion-based generative modeling, and global optimization into a unified methodology. First, microstructure images are transformed into a compact "morphology fingerprint" that captures multi-scale texture, particle sphericity, and size distribution metrics via image based quantitative morphology analysis.[9, 10] Subsequently, a conditional diffusion model functions as a forward simulator, synthesizing realistic SEM images from process parameters or target descriptors with high fidelity.[11, 12] Next, a Particle Swarm Optimization (PSO) algorithm is employed to iteratively adjust coprecipitation conditions.[13] This process is intended to direct the generated images toward the desired morphology. The efficacy of each iteration is evaluated using quantitative morphology metrics, which serve to determine the algorithm's "fitness." By treating the microstructure image itself as the design objective rather than relying on scalar proxies, our approach actively explores the space of possible morphologies in a feedback loop. We validated this framework by synthesizing Li- and Mn-rich layered oxide cathodes under the optimized conditions predicted by our algorithm, and SEM characterization confirmed that the experimentally obtained microstructures closely matched the target morphology fingerprint.

# Integrated Framework for Forward Prediction and Inverse Design

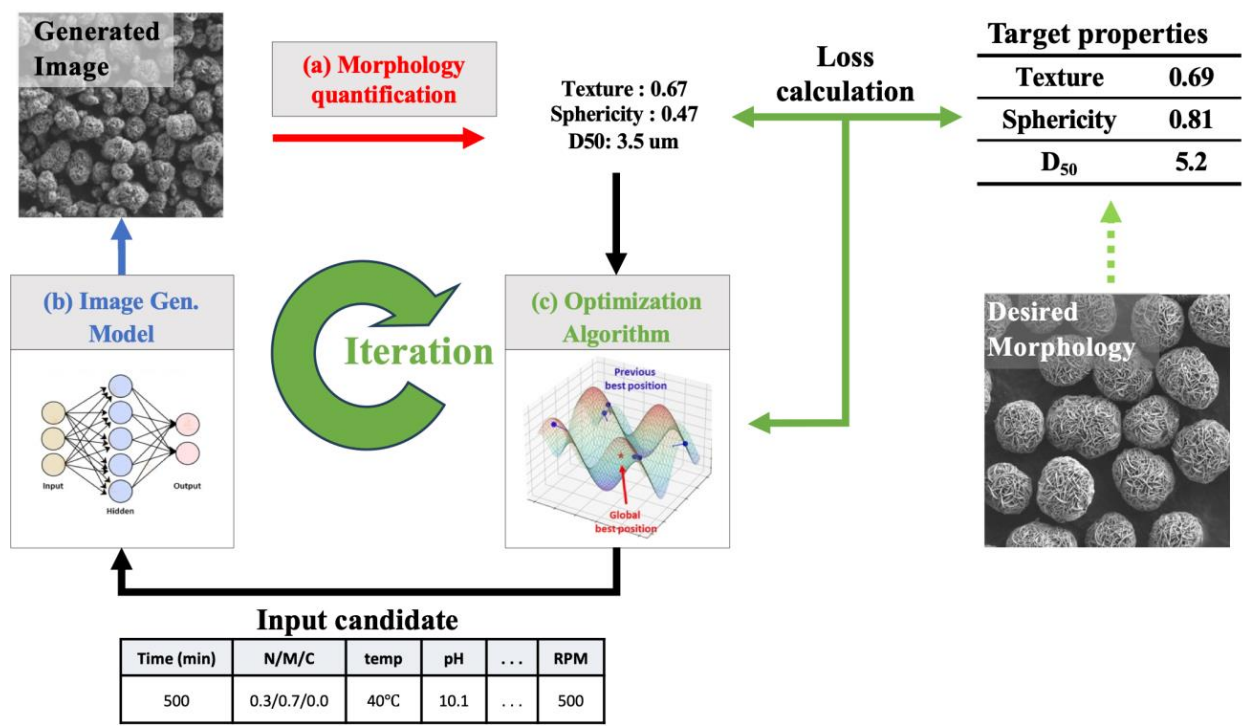

**Figure 1. Schematic overview of the integrated AI-based framework for precursor morphology optimization.** The framework consists of three key components: (a) a morphology analysis module that quantifies texture, sphericity and $D_{50}$ from the images, (b) a diffusion-based image generation model that predicts SEM-like precursor morphologies from coprecipitation parameters, and (c) a particle swarm optimization (PSO) algorithm that iteratively refines synthesis conditions to minimize the difference between predicted and target morphology. Given a user-defined target morphology(right), the framework iteratively generates precursor images under candidate synthesis conditions, quantifies their morphology features, and updates the input parameters until convergence is reached. The bottom table shows an example of the coprecipitation parameters used as model input.

The proposed workflow (Figure 1) reimagines precursor synthesis as a closed-loop, data-driven process, rather than the traditional, time-intensive trial-and-error approach. The system is composed of three synergistic modules that form the backbone of the overall structure. First, a morphology analysis model was developed that extracts three reproducible descriptors directly from SEM micrographs. The developed model supplants qualitative labels such as "needle-like" or "plate-like" with a texture metric,[14-16] and it replaces conventional laser-diffraction PSA measurements with sphericity and particle-size metrics, thereby providing richer, SEM-scale insights. Secondly, a conditional diffusion-based image generator has been developed to learn the mapping between coprecipitation parameters (*i.e.*, pH, NaOH/NH$_4$OH ratio, reaction time) and realistic SEM-like morphologies. This model enables the reconstruction of multiscale morphological features, thereby facilitating rapid virtual screening of admissible parameter sets.[11, 12, 17, 18] Thirdly, a particle swarm optimization (PSO) routine interrogates the generator iteratively to converge on synthesis conditions that meet user-defined morphological targets.[13] PSO ultimately completes the cycle by sampling the design space, querying the generator for a virtual image, extracting descriptors through the analysis model, and updating its velocity based on the discrepancy between its current state and the target morphology.

## Image Based- Quantitative Morphology Analysis

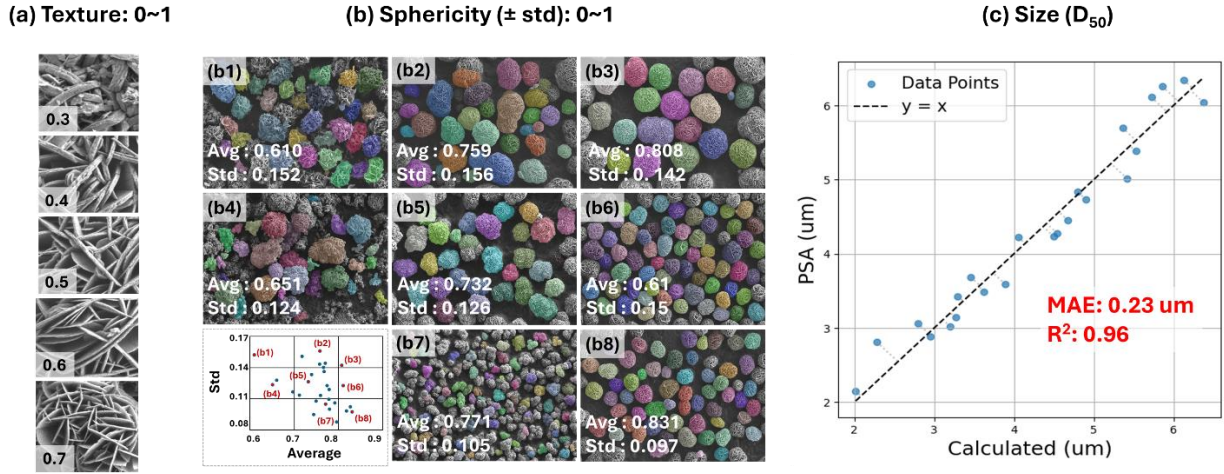

**Figure 2. Quantitative image-based analysis of precursor morphology using texture, sphericity, and $D_{50}$.** **(a)** Texture quantification based on discrete wavelet transform (DWT),[10] used to distinguish fine primary particle structures reflecting differences in surface microstructural complexity. **(b)** Sphericity analysis of secondary particles visualized with color-coded segmentation. Average sphericity (Avg) and standard deviation (Std) values are presented for each sample (bottom-left graph) **(c)** Comparison between $D_{50}$ values extracted from image analysis and experimental particle size analyzer measurements across different synthesis times. The close match validates the reliability of the image-based quantification method.

Domain-knowledge based morphology descriptors are indispensable for correlating precursor microstructure with electrochemical and mechanical behavior. Accordingly, this study employs the two metrics already standard in powder engineering the median equivalent diameter ($D_{50}$) and the sphericity of secondary particles (agglomerates) reporting both their population means and full distributions.[19-23] To complement these metrics, we introduce a texture descriptor that quantitatively captures primary-particle shape, which prior qualitative labels could not describe consistently.

Sphericity and $D_{50}$ of secondary particle are widely used because they predict packing density, stress distribution during electrode calendaring, and fracture resistance under compression.[24] AI assisted segmentation enabled calculation of both $D_{50}$ and sphericity from projected area and equivalent radius. (Figure 2b) The resulting $D_{50}$ values matched laser diffraction data within ±4 % (Figure 2c).

The texture descriptor fills the gap left by subjective terms such as "plate-like", "road-like" or "needle-like."[14-16] Although such qualitative labels may hint at the dominant crystallographic facets, they are inherently unreliable because their meaning is observer-dependent and they reduce complex three-dimensional geometry to a single adjective. Most importantly, they do not provide a continuous metric that can be incorporated into engineering models or statistical analyses.

Because texture is computed directly from voxel-level geometry, it remains reproducible even when phase boundaries are indistinct, enabling robust quantification of primary particle shape (Figure 2a). Several groups have attempted to quantify primary-particle shape by first segmenting the particles and then extracting geometric metrics from the segmented volumes. While conceptually attractive, this "direct segmentation" route becomes highly sensitive to hyper-parameter choices or to the specific distribution of labels in the training set once inter-phase boundaries are diffuse or crystallographic domains intergrow.

To verify that the selected descriptors carry complementary information, we calculated pairwise Pearson correlation coefficients among $D_{50}$, mean sphericity, and texture (Supplementary Figure S4). The values were −0.38 ($D_{50}$ vs. sphericity), −0.17 ($D_{50}$ vs. texture), and 0.26 (sphericity vs. texture), indicating only weak correlations. This confirms that the domain-knowledge-based descriptor set is mutually independent and therefore suitable for quantitative morphology analysis.

Correlation with experimental variables further shows how particle shape can be controlled. Texture exhibited strong dependence on pH, initial NaOH/NH$_4$OH ratio, and synthesis time, rising to ≥ 0.7 under an initial NaOH/NH$_4$OH ratio of 3.33 at pH 10 (thin, faceted lamellae) and falling below 0.3 when the ratio dropped to 0.15 at pH 11 or when O$_2$-induced Mn$^{2+}$ oxidation promoted platelet thickening. Sphericity increased monotonically with reaction time while its standard deviation decreased, indicating progressive homogenization; higher NaOH/NH$_4$OH ratios enhanced nucleation density and yielded more spherical particles, whereas lower pH or oxygen gas reduced sphericity.

# Forward Prediction via Diffusion-Based Image Generation

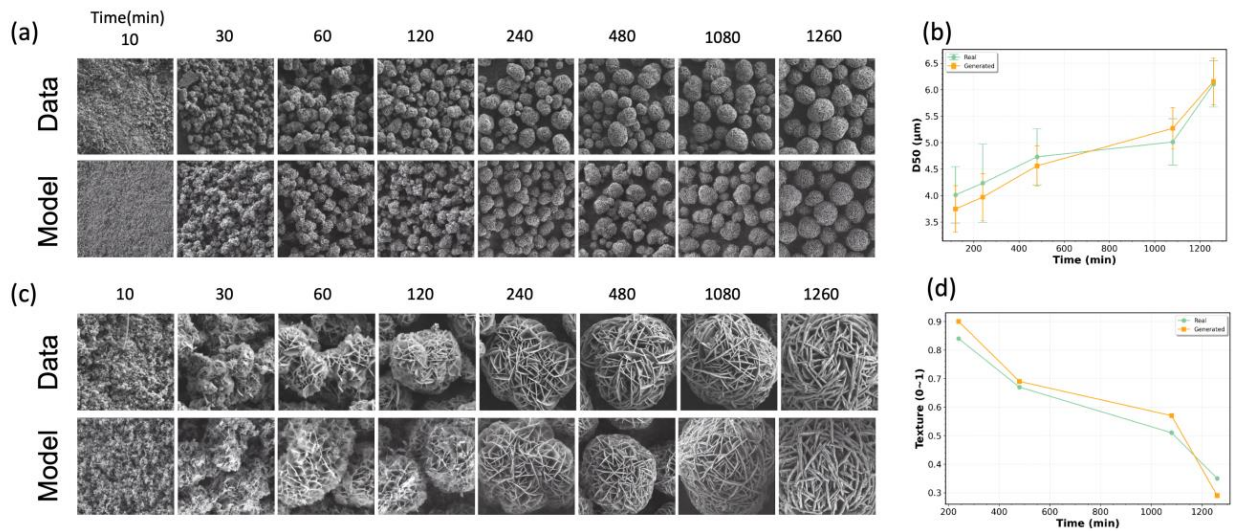

**Figure 3. Comparison of experimental and model-generated precursor morphologies and their quantitative descriptors across synthesis time. (a)** Side-by-side comparison of experimental SEM images (top row) and diffusion model-generated images (bottom row) across different synthesis durations at high magnification (15,000×), highlighting the evolution of primary particle texture. **(b)** Quantitative comparison of texture values extracted from experimental and generated images, showing consistent trends over synthesis time. **(c)** Corresponding comparison at lower magnification (3,000×), capturing the development of secondary particle morphology and packing structure. **(d)** Comparison of $D_{50}$ values (median secondary particle size) between experimental SEM images and model-generated image-based estimations. The close agreement in both metrics supports the validity of the generative model and the reliability of the image-based analysis framework.

The conditional diffusion model translates arbitrary coprecipitation parameters into realistic SEM-like micrographs in <5 s on a consumer GPU. Figure 3a juxtaposes experimental and generated 15,000× images, highlighting accurate replication of the rapid texture relaxation observed- even in the very first nucleation steps (~ 120 min). At 3,000× images (Figure 3c), secondary particle densification and aggregate coalescence are reproduced with striking visual fidelity.

Additionally, although the model does not rely on explicit physical equations, it learns the relationship between coprecipitation conditions (pH, NaOH, NH$_4$OH, time, etc.) and precursor morphology, generating results that mirror the actual growth process. In real coprecipitation, initial nuclei rapidly aggregate into agglomerates, provoking swift morphological change; thereafter, crystals inside the agglomerates grow more slowly.[25, 26] The diffusion model successfully reproduces this behavior: during the first 120 min it captures the fast evolution of secondary and primary particles, and at later times it recreates the Ostwald-ripening tendency whereby convex regions dissolve, concave regions grow, and overall particle sphericity increases.[27] Quantitatively, texture values derived from generated images track experiments with $R^2 = 0.94$ (Figure 3b); D$_{50}$ trajectories align with $R^2 = 0.91$ (Figure 3d), demonstrating- that the model preserves underlying structural statistics rather than merely mimicking style.

Our model demonstrates robust performance within the training manifold, effectively capturing morphology variations across multiple synthesis parameters. As shown in Supplementary Figure S5, morphology evolution under varying pH conditions is accurately predicted. Notably, the model successfully generates realistic morphologies even at interpolated conditions, such as pH 10.5, which were absent from the training dataset. This result highlights the model's capability for generalization and its alignment with known physical growth mechanisms, including directional adsorption of metal-ammonia complexes on the (001) plane

Building upon this capability, Supplementary Figure S6 demonstrates forward predictions in a two-dimensional synthesis parameter space, wherein both pH and initial ammonia concentration are simultaneously varied. Training data were provided exclusively at the boundary conditions of 1.0 wt% NH$_4$OH at pH 10.0 and 2.0 wt% NH$_4$OH at pH 11.0, with intermediate conditions interpolated at increments of 0.2. The results show a clear decreasing trend in $D_{50}$ values as pH and ammonia concentration increase, while sphericity remains relatively constant across the interpolation space.

# Inverse Design- Validation through PSO Optimization

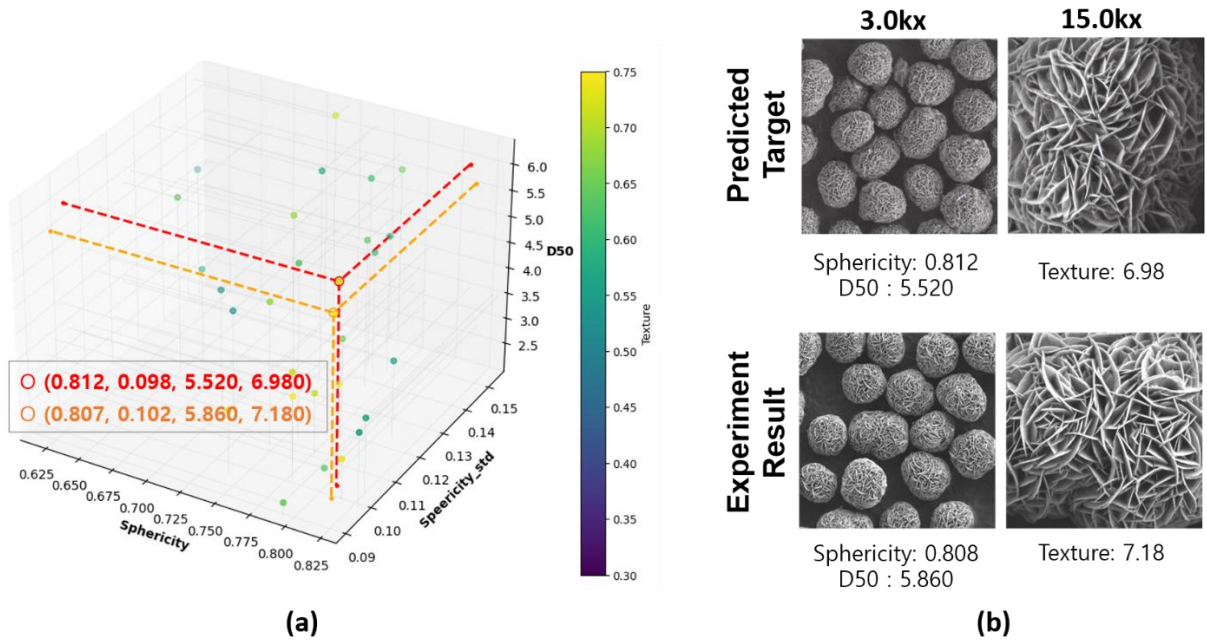

**Figure 4. Optimization of precursor morphology using an integrated image-based inverse design framework. (a)** Visualization of the optimization result in a three-dimensional morphological design space defined by sphericity, sphericity standard deviation, and $D_{50}$ (median secondary particle size). The color scale represents texture, with warmer colors indicating finer primary features. Each point corresponds to an experimentally measured dataset. The red star marks the predefined target morphology, and the blue star denotes the optimized result. **(b)** SEM image comparison of the predicted target morphology (top row) and the experimentally synthesized result (bottom row) obtained using the optimized coprecipitation conditions. Low and high magnification images are shown to evaluate both secondary particle shape and primary particle texture.

To showcase practical utility, we defined a target morphology characterized by high texture, high sphericity, low σ(sphericity), and elevated $D_{50}$—traits associated with fast Li$^+$ diffusion and high tap density. PSO launched 40 search points across the feasible parameter space, each iteratively querying the generator, quantifying descriptors, and updating its trajectory. Convergence occurred within 50 iterations, yielding the following optimal synthesis conditions: sealed reactor condition enabled, initial NaOH concentration of 0.3 wt%, initial $NH_4OH$ concentration of 0.3 wt%, and additive content of 4000 ppm (Figure 4a).

Experimental validation confirmed predictive accuracy. SEM images of the synthesized precursor (Figure 4b, bottom) reveal densely packed secondary particles and finely textured primary networks that mirror virtual predictions (top). Descriptor discrepancies between prediction and experiment remain ≤5 % across all metrics.

## Discussion

These findings confirm that the proposed inverse-design approach effectively guides experimental synthesis toward target morphological features, underscoring its potential for rational microstructure control in materials design. By replacing vague descriptors with a triad of quantitative metrics texture, sphericity, and $D_{50}$ the SEM image itself becomes a first-class design variable. Correlation analyses (*Image Based Quantitative Morphology Analysis* section) revealed that each descriptor responds independently to pH, NaOH/NH$_4$OH ratio, and reaction time, thereby delineating a controllable processing window for tap-density and Li-ion-diffusivity and capacity retention. These correlations clarify the specific operating space needed to tailor secondary-particle morphology to target electro-mechanical performance (Supplementary Figures S2, S3), a link that coarse surrogates such as BET or PSD previously obscured.

The high-fidelity synthetic micrographs generated entirely in silico demonstrate the feasibility of rapid morphology scouting across extensive compositional spaces (e.g., the full pH–NH$_4$OH range) in mere minutes, thereby compressing weeks of wet-lab trial-and-error into GPU time. Every pixel in the conditional diffusion generator is conditioned on the same descriptor set used for experimental quantification, allowing researchers to trace visual motifs back to numeric targets. Tens of thousands of hypothetical conditions can thus be screened per hour on a single consumer GPU—throughput unattainable with conventional CSTR campaigns—and the linear scaling of optimization cost with descriptor count permits seamless integration of further constraints (impurity limits, BET, grain-boundary texture) without architectural overhaul.

Some localized deviations from monotonic behavior observed in interpolated morphologies (*Forward Prediction via Diffusion-Based Image Generation* section) can be attributed to nonlinear interactions between chemical equilibria (e.g., complexation efficiency, supersaturation) and kinetic factors (e.g., nucleation, coarsening). Such discrepancies are common in multivariate synthesis landscapes and highlight opportunities for refinement through expanded data coverage or conditioning guided by physical constraints.

Although demonstrated on Li- and Mn-rich layered-oxide precursors, the framework is inherently material agnostic: swapping the training image set and descriptor definitions would enable immediate transfer to catalysts, additively manufactured alloys, or porous membranes where geometry governs function. By translating explicit morphological targets into experimentally verifiable recipes, the integrated AI system furnishes a blueprint for accelerating morphology-driven discovery not only in battery materials but also in catalytic, pharmaceutical, and structural domains.

Several challenges remain before the full implementation of autonomous morphology design becomes a standard practice.

- **Data-domain breadth:** The present model's training is limited to a single reactor platform and a limited pH–temperature domain; federated learning across multiple synthesis routes will improve domain generality and robustness against covariate shift.

- **Property-level integration:** Electrochemical performances are still analyzed offline. The incorporation of cycling data directly into the fitness function would facilitate a unidirectional loop that couples structure and property optimization.
- **Non-linear interaction capture:** Augmenting the training set around regions that exhibit complex kinetics–thermodynamics coupling, or applying physics-informed priors, should mitigate residual error.

The ongoing work integrates multi-reactor datasets and streams real-time electrochemical feedback, with all code and pretrained weights to be released under an open-source license. Coupling this image-driven inverse-design paradigm with robotic synthesis and high-throughput imaging holds great promise for achieving truly closed-loop optimization, thereby accelerating the discovery process for lithium-ion batteries and beyond. We anticipate that data-guided morphology engineering will soon advance from proof-of-concept to a standard tool in the materials-by-design platform, ultimately accelerating the decades-long lab-to-fab timeline for morphology-sensitive materials.

# Image-Guided Microstructure Optimization using Diffusion Models: Validated with Li-Mn-rich Cathode Precursors

Geunho Choi,[‡] Changhwan Lee,[‡] Jieun Kim, Insoo Ye, Keeyoung Jung[*] and Inchul Park[*]

**Supplementary Information**

## 1. Experimental methods

We introduce an integrated computational framework designed to predict optimized synthesis parameters for achieving target particle morphologies. The framework architecture comprises three interdependent modules: (1) a morphology quantification algorithm that systematically analyzes particle characteristics, including primary and secondary particle shapes from microscopy images; (2) a diffusion-based generative framework incorporating ControlNet conditioning mechanisms to synthesize physically consistent particle morphology images constrained by user-specified synthesis conditions; and (3) an inverse prediction algorithm that leverages these two components to determine optimal synthesis conditions for target morphologies. This comprehensive pipeline enables both forward prediction of morphological outcomes and inverse design of synthesis parameters. The subsequent sections delineate each component in detail, beginning with the development of the quantitative morphology analysis pipeline, followed by the implementation of the ControlNet-guided diffusion model for conditional image synthesis, and concluding with the inverse design framework that bridges synthetic controllability with morphological predictability. The integration of diffusion-based generation with explicit control mechanisms ensures high-fidelity representation of synthesis-structure correlations while maintaining physical plausibility across parameter spaces.

### 1.1 Data Preparation

For the training of the image generation model, we aimed to obtain a minimal training dataset through a precursor coprecipitation experiment. The precursor synthesis was conducted using a water-jacketed continuous stirred-tank reactor system. A Mn-rich composition (Ni:Mn = 33:67) was synthesized using $MnSO_4 \cdot H_2O$ and $NiSO_4 \cdot 6H_2O$ (2.0 M total), with NaOH (10.0 M) and $NH_4OH$ (7.7 M) added to maintain pH values of 9.8, 10.0, 10.7, or 11. Initial co-precipitation conditions were adjusted by pre-adding NaOH and $NH_3$ to the reactor. The synthesis was carried out at 30 °C under continuous stirring (1000 rpm) and nitrogen atmosphere. The final product was washed, filtered, and vacuum-dried at 100 °C overnight.

The synthesized precursor was examined using a scanning electron microscope (SEM; JCM 6000, JEOL) equipped with a conventional secondary electron detector to acquire high-resolution images of its structural features. The imaging process was performed under high vacuum conditions with an accelerating voltage of 2 kV to ensure minimal charging effects and optimal resolution. To systematically evaluate the hierarchical morphology of the precursor, two distinct magnification levels 3.0k and 15.0k were employed. The lower magnification (3.0k) facilitated a macroscopic assessment of the secondary particles, enabling observation of their overall size distribution, aggregation behavior, and surface morphology. In contrast, the higher magnification (15.0k) provided a nanoscale resolution to scrutinize the fine structural details of the primary particles, including their shape, crystallinity, and interfacial boundaries. The obtained images were utilized as a dataset for training a image generation model, as detailed in subsequent sections, enabling systematic prediction of quantitative morphological variations under varying synthetic parameters. The particle size distribution of the synthesized precursors was measured using a particle size analyzer (Microtrac S3500), providing $D_{50}$ values.

### 1.2. Morphology Quantification via Digital image processing

The morphological characteristics of LLO particles play a crucial role in determining electrochemical performance, as extensively documented in the literature.[1, 2] Traditional analytical approaches have predominantly relied on qualitative visual assessment of features such as primary particle aspect ratios, plate-like, and needle-like formations, along with secondary particle characteristics including aggregation states, fragmentation patterns, and overall sphericity.[3-5] However, such subjective evaluations present significant limitations for comparative analyses and systematic optimization. To address these limitations and enable integration with our image generation framework, we developed quantitative methodologies for characterizing both primary and secondary particle morphologies.

### 1.2.1 Quantitative Analysis of Primary Particle Morphology via Texture Analysis

Accurate morphological characterization of primary particles in cathode materials is critical for correlating microstructure with electrochemical performance. However, direct quantification through individual particle segmentation often considered the gold standard for aspect ratio and shape analysis faces intractable challenges in practical applications. High-magnification SEM images frequently exhibit complex particle overlap, ambiguous boundaries, and entanglement, rendering conventional segmentation methods error-prone and unreliable. To address these limitations, we developed an indirect yet robust texture analysis framework to quantify morphological features without requiring explicit particle isolation. The proposed methodology involves a multi-step computational pipeline applied to 15.0k SEM images. First, background noise and non-particle artifacts are systematically eliminated to enhance feature specificity. Automated detection of particle centers is then performed, enabling the extraction of 256x256 pixel regions of interest (ROIs) centered on individual particles. To mitigate contrast variability inherent to SEM imaging, all ROIs undergo histogram equalization, ensuring consistent intensity distributions across the dataset. Subsequent texture characterization leverages wavelet decomposition (via PyWavelets) to disentangle hierarchical structural patterns within each ROI.[6] This approach captures both coarse and fine-grained textural signatures, which correlate with morphological attributes such as surface roughness, edge sharpness, and particle anisotropy.[7] The derived wavelet sub-bands are quantified using an energy metric defined as:

$$\text{Energy} = (1/(N \times M)) * \Sigma \Sigma |W(i,j)|^2 \quad (1)$$

where $W(i,j)$ represents the wavelet coefficient at position $(i,j)$, and $N \times M$ denotes the sub-band dimensions. This scalar metric encapsulates the textural complexity of each particle, effectively serving as a proxy for its morphological characteristics. By aggregating energy values across multiple decomposition levels and orientations, the method achieves a comprehensive representation of particle morphology while inherently accommodating overlap and boundary ambiguities.

This texture-centric strategy circumvents the pitfalls of direct segmentation, offering a reproducible and contrast-invariant framework for quantifying primary particle morphology in densely aggregated cathode materials. The energy metric's sensitivity to structural heterogeneity further enables statistical comparisons across synthetic batches or processing conditions, providing actionable insights for microstructure optimization.

### 1.2.2 Quantitative Analysis of Secondary Particle morphology via Instance Segmentation

To quantify the sphericity of secondary particles in low magnification scanning electron microscopy (SEM) images of cathode materials, precise instance segmentation of individual particles was required. We applied the Segment Anything Model (SAM)[8], a zero-shot deep learning segmentation framework, to ensure accurate and reproducible particle boundary identification. The segmentation protocol included automated exclusion criteria to remove particles intersecting image boundaries or exhibiting overlap, thereby restricting analysis to fully isolated particles with intact morphological features. Sphericity was quantified from the segmented particle masks using a circularity-derived metric defined as:

$$\Psi = 4\pi A/P^2 \quad (2)$$

where $\Psi$ denotes the sphericity index (ranging from 0 to 1), A is the projected area of the particle, and P is its perimeter. A value of $\Psi=1$ indicates a perfectly circular projection, while decreasing values reflect progressive deviation from spherical morphology. This dimensionless index provides statistically relevant measures of particle uniformity and aggregation behavior within composite cathode architectures.

The methodology enables rigorous batch-level morphological analysis while resolving intrinsic interparticle heterogeneity. By coupling single-particle sphericity measurements with ensemble-level statistical descriptors, this multiscale quantification framework establishes numerically grounded correlations between particle morphology and synthesis-process parameters, serving as critical inputs for inverse design optimization.

### 1.2.3 Synthetic image generation

Diffusion models are generative frameworks grounded in the mathematical simulation of stochastic diffusion processes, operating through two interconnected phases: forward and reverse diffusion.[9] In the forward phase, input data such as images or material microstructures is progressively perturbed by injecting Gaussian noise over a predefined sequence of timesteps. This systematic corruption mimics thermodynamic diffusion, gradually transforming structured data into isotropic Gaussian noise. The reverse phase, governed by probabilistic principles, learns to invert this entropy-driven degradation by approximating the score function of the data distribution. Through iterative denoising, the model reconstructs the original structure, enabling high-fidelity sample generation with improved stability and mode coverage compared to adversarial methods. This formalism provides a robust mathematical foundation for capturing complex, high-dimensional data distributions, making diffusion models particularly effective for applications ranging from image synthesis to materials design.

To address the computational inefficiencies of pixel-space diffusion, Stable Diffusion repositions the generative process into a compressed latent space using a variational autoencoder (VAE).[10, 11] The VAE encodes high-dimensional data into a lower-dimensional latent representation, preserving critical structural features while significantly reducing computational overhead analogous to projecting multiphase microstructures into reduced-order descriptors. By executing diffusion in this compressed domain, Stable Diffusion achieves substantial gains in memory efficiency and processing speed without sacrificing output quality.

Furthermore, the framework incorporates cross-modal conditioning through text embeddings derived from contrastive language-image pretraining (CLIP) models.[12] Trained on the LAION-SB dataset,[13] which contains vast image-text pairs, Stable Diffusion learns to associate natural language prompts (e.g., "nanoporous thin film with 20 nm pore spacing") with corresponding visual or structural patterns. This capability supports diverse applications, including text-guided microstructure generation and defect-aware image inpainting, establishing the model as a versatile tool for multiscale synthesis tasks.

Despite its strengths in open-ended generation, Stable Diffusion struggles to enforce domain-specific constraints essential for materials science applications, such as crystallographic orientations or spatially resolved phase distributions. To overcome this limitation, ControlNet is introduced as an auxiliary architecture designed to impose precise structural or semantic guidance without altering the pretrained Stable Diffusion model.[14] ControlNet duplicates the encoder layers of the base model and integrates zero-initialized convolutional layers trained to process auxiliary conditioning signals. These signals-which may include synthesis parameters like temperature gradients, stoichiometric ratios, or topological constraints specific to battery cathode materials are encoded as edge maps, compositional gradients, or other domain-specific descriptors. Crucially, the pretrained weights of Stable Diffusion remain frozen during training, preserving its generative diversity and preventing catastrophic forgetting. The zero-initialized layers enable incremental integration of conditioning features, stabilizing training dynamics while learning intricate correlations between input constraints and output structures.

The adoption of ControlNet addresses a critical challenge in conventional diffusion models: balancing precise control with generative flexibility. By decoupling conditional guidance from the base generative process, the architecture allows users to steer synthesis toward specific objectives such as spherical secondary particles or needle-like shaped, dense primary particle arrangements without retraining the base model. The normalized synthesis conditions were combined into a single conditioning vector and fed into the ControlNet architecture. This architecture consists of three hidden layers with ReLU activation functions, designed to align with the dimensionality of the diffusion model's feature representation. Throughout the training process, the base diffusion model remained frozen, while the ControlNet was optimized using an extensive dataset of synthesized materials paired with their corresponding synthesis conditions. This approach ensured that the ControlNet could effectively learn and adapt to the relationships between the synthesis parameters and the resulting material properties

### 1.2.4 Inverse prediction framework for optimal synthesis conditions

This study proposes an inverse design framework to determine optimal synthesis parameters for achieving target particle morphologies. The methodology integrates a ControlNet-based image generation model, a morphology quantification algorithm, and Particle Swarm Optimization (PSO) to systematically navigate the synthesis parameter space.[15] Users may specify target morphological characteristics such as sphericity and surface texture either as numerical values or reference images, and the framework autonomously identifies experimental conditions that satisfy these specifications.

This optimization process begins by generating an initial population of candidate synthesis parameters randomly distributed within experimentally feasible bounds. For each candidate parameter set, the ControlNet model synthesizes corresponding particle images by leveraging its pretrained understanding of the relationship between synthesis conditions and morphological outcomes. These generated images are then analyzed using a quantitative morphology algorithm to extract numerical descriptors of size ($D_{50}$), sphericity (M), and texture (T). The discrepancy between the generated and target characteristics is evaluated through a weighted objective function:

$$F = w_1\sum(D_{target} - D_{generated})^2 + w_2\sum(M_{target} - M_{generated})^2 + w_3\sum(T_{target} - T_{generated})^2 \quad (3)$$

where $w_1$, $w_2$ and $w_3$ are empirically derived weighting coefficients that balance the relative importance of morphological and textural fidelity. Guided by PSO, the framework iteratively refines the candidate parameters by combining swarm-based exploration of the parameter space with convergence toward solutions that minimize the objective function. This iterative process continues until predefined convergence criteria are met, yielding synthesis conditions that optimally replicate the target morphology. The multiple-input capability accepting both numerical targets and reference images enhances the framework's versatility, making it applicable to both data-driven and experimentally grounded materials design scenarios.

## 2. Texture Quantification Using DWT

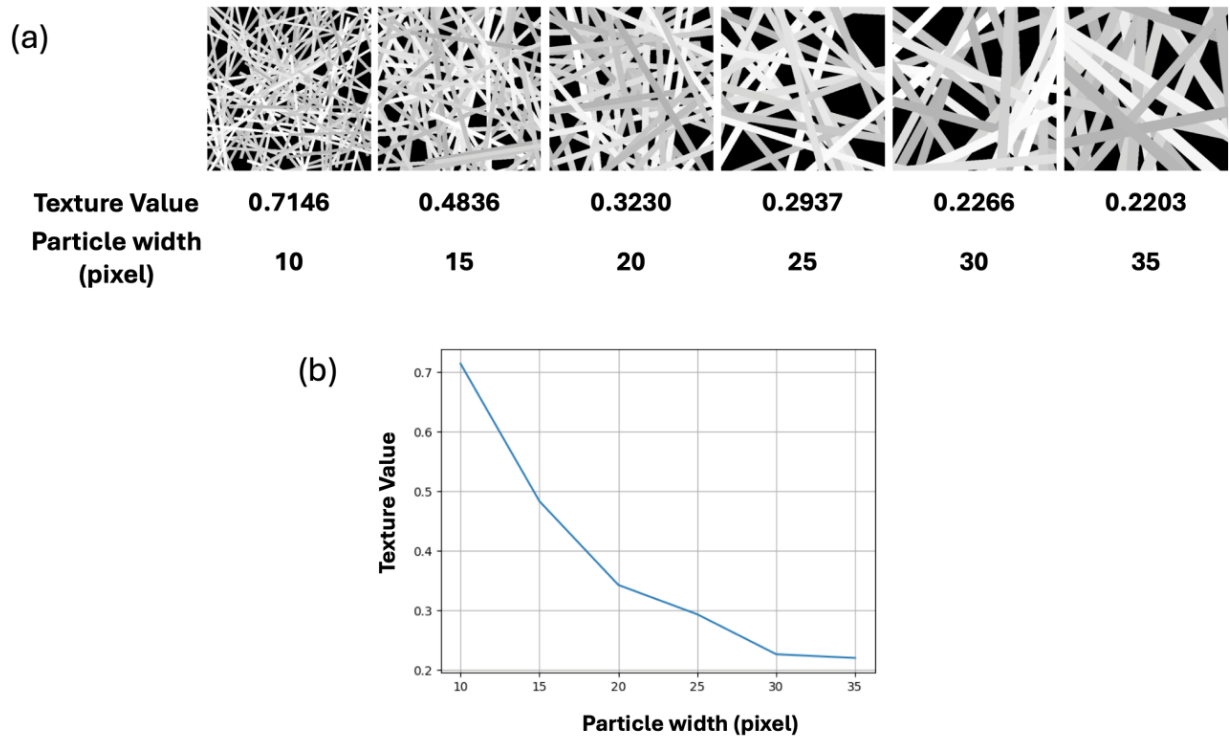

**Figure S1. Simulated images of particle networks and their corresponding texture values extracted using Discrete Wavelet Transform (DWT).** **(a)** As particle width increases, the texture value decreases due to reductions in high-frequency components, pattern density and spatial complexity. **(b)** The line plot visualizing the invers correlation between particle width and DWT-derived texture value, highlighting the method's sensitivity to morphological variations.

To quantitively assess the textural complexity of precursor, we employed a Discrete Wavelet Transform (DWT) to extract texture values from simulated images with varying particle widths. As shown in Figure S1, the texture value exhibits a decreasing trend as particle width increases. This behavior can be attributed to four primary factors. First, images with smaller particle widths contain finer structural features and more intricate local patterns, resulting in a higher proportion of high-frequency components in the wavelet domain. Second, these finer structure exhibit high pattern density, where numerous overlapping elements increase local variability and visual complexity. Third, spatial complexity is also greater in images with thinner particles, as the randomness and irregularity of interesting structure are more pronounced. In contrast, images with wider particles present more homogeneous and repetitive patterns, leading to reduced textural richness. Lastly, the energy distribution of wavelet coefficients plays a critical role: texture values are elevated when a significant portion of the energy resides in the high-frequency bands, which is characteristic of more complex and finely detailed images. These combined effects confirm that the DWT-based metric effectively captures the structural richness of images and sensitivity reflects morphological variation in particle networks.

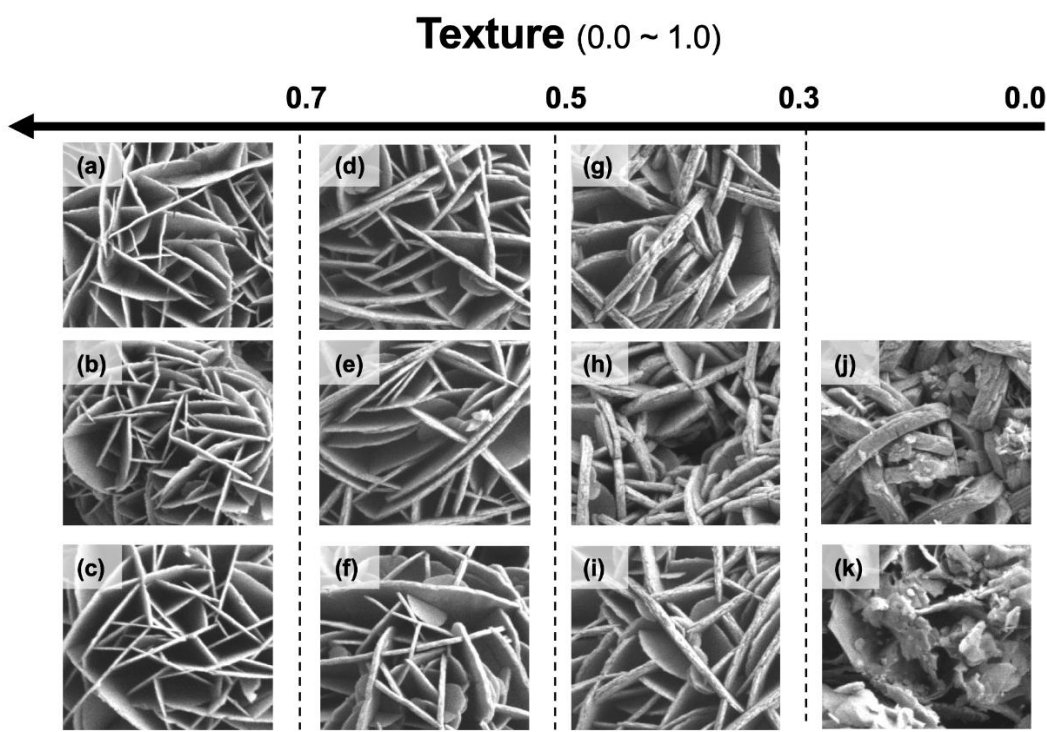

**Figure S2. Representative SEM images categorized by normalized texture value for primary particle analysis.** Cropped SEM images (2.0k magnification) grouped into four texture ranges: **(a–c)** high texture (1.0–0.7), **(d–f)** medium-high texture (0.7–0.5), **(g–i)** medium-low texture (0.5–0.3), and **(j–k)** low texture (0.3–0.0). As the texture value decreases, primary particles become thicker, less defined, and increasingly aggregated. This classification highlights the morphological diversity captured by the texture quantification model.

To further validated the applicability of the texture quantification approach, Figure S2 presents a captured set of SEM images categorized by texture values. This textures matric, derived using

DWT, captures fine-scale morphological complexity by evaluating high-frequency content, pattern density, and spatial irregularity. Images were grouped into four texture ranges for comparative analysis: high (1.0-0.7), medium-high (0.7-0.5), medium-low (0.5-0.3), and low (0.3-0.0).

Samples (a-c) exhibit the high texture values, featuring thin, well-defined primary particle and high surface intricacy. These characteristics typically emerge from co-precipitation conditions with a high initial $NaOH/NH_4OH$ mol ratio of 3.33 and a pH 10, which suppress surface adsorption of complex ions and slow the growth of primary particle, maintaining fine structures.[4, 5] In contrast, sample (d-f) and (g-i) display progressively reduced texture values, with primary particles becoming ticker, more stacked, and less geometrically defined. Samples (a), (d) and (g) were obtained under identical synthesis conditions but at increasing reaction times, illustrating time-dependent coarsening behavior and corresponding decline in texture.[16] In the lowest texture group (j-k), morphological degradation becomes evident. Sample (j), synthesized at a low initial $NaOH/NH_4OH$ mol ratio of 0.15 and a pH 11, shows wide, aggregated plates due to enhanced deprotonation at the precursor surface resulted in promoted preferential growth along the [001] direction.[3] Meanwhile, sample (k), demonstrates severe textural collapse induced the presence of O2 during co-precipitation, where $Mn^{2+}$ oxidation disrupted crystal growth, yielding fragmented aggregates.[17, 18] This classification highlights the ability of DWT-based texture metric to distinguish subtle yet significant variations in precursor morphology, providing a robust foundation for subsequent modeling and optimization.

## 3. Sphericity-Based Classification of Secondary Particle Morphology

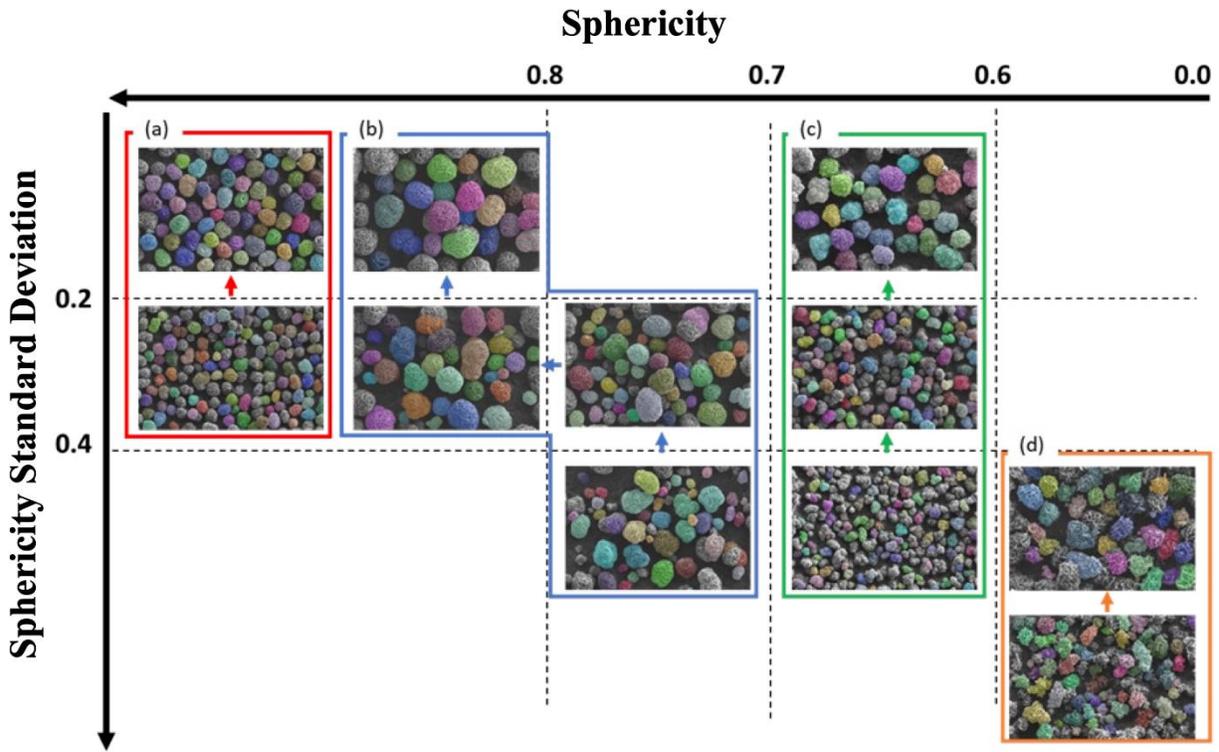

**Figure S3. Representative SEM images showing the evolution of secondary particle sphericity.** Each group of images (a–d) corresponds to a distinct coprecipitation condition, with time increasing from top to bottom. All images are shown at 3.0k× magnification, with colorized segmentation to highlight individual particles. **(a)** Highly spherical particles with low variance. **(b)** Well-rounded particles with moderate uniformity. **(c)** Developing sphericity with shape heterogeneity. **(d)** Irregular, distorted particles with low sphericity and high variance. The sequence illustrates how synthesis time and solution parameters influence the development of particle roundness and packing uniformity.

The average sphericity and variance of secondary particles in precursor images were calculated, and the precursor images were organized according to Figure S3. The precursor groups in the upper-left quadrant, which have high average sphericity and low variance, also exhibit nearly spherical particle shapes and uniform morphology in actual images. In contrast, precursor groups in the lower-right quadrant, which have low sphericity and high variance, display individual or distorted particle shapes. The classified precursor groups generally exhibit a trend where variance in sphericity decreases over time. While there are coprecipitation conditions that increase sphericity, most conditions maintain a stable sphericity without significant increases. Figure S3a and 3b show samples where all conditions are identical except for the initial NaOH/NH$_4$OH mol ratio, which is 3.33 and 0.33, respectively, allowing an observation of how the initial solution influences sphericity. High initial NaOH concentration increases the nucleation density in the solution. The higher nucleation density reduces the diffusion coefficient of particles, preventing easy aggregation and instead forming uniformly distributed spherical particles through surface diffusion, as previously reported. In contrast, in the low NaOH concentration sample, the particles initially exhibited low and irregular sphericity due to aggregation, but over time, the particles transitioned to a more uniform and higher sphericity.[19] Figure S3c presents two samples under identical conditions as Figure S3a, except for different pH levels of 10.7 and 10. Similar to the previous texture results, an increase in pH enhances metal-ammonia complex adsorption, leading to dominant primary particle growth, which subsequently affects the sphericity of secondary particles.[3] Figure S3d represents precursors grown in an oxygen-rich atmosphere, where oxidation of the particle surface resulted in a distorted shape. In this case, particles aggregated during the intermediate reaction stage but did not undergo re-dissolution and re-precipitation, preventing them from forming a spherical shape.[17, 18] This study confirms that sphericity can serve as a systematic and quantifiable metric for characterizing precursor morphology, further supporting its use in optimizing coprecipitation conditions.

## 4. Mutual Independence of Sphericity, Texture, and Size

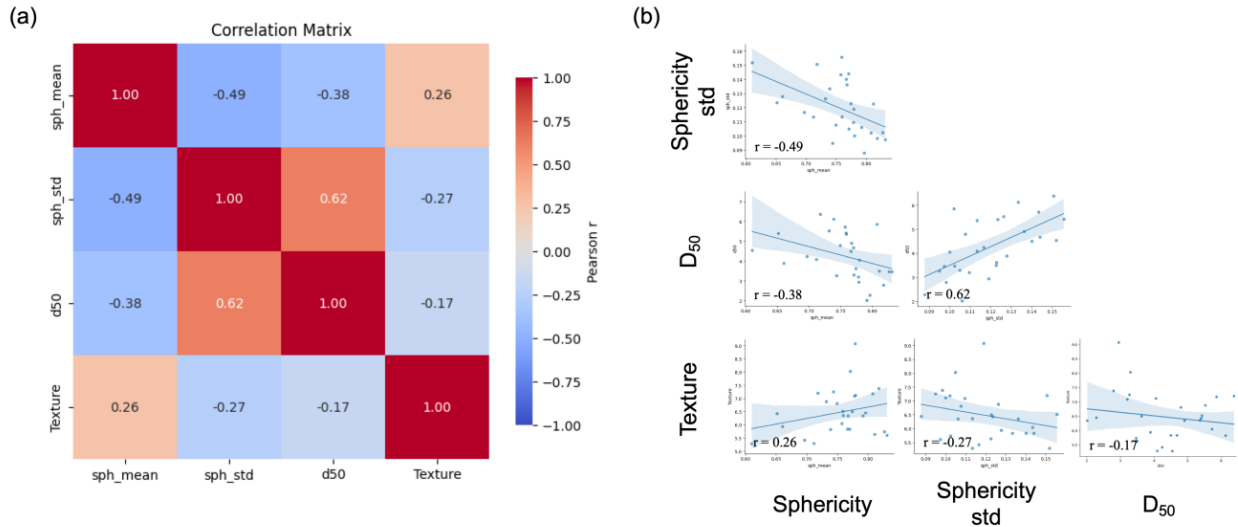

**Figure S4. Quantitative correlation among morphology descriptors.** (a) Pearson correlation matrix among four morphology descriptors: mean sphericity (sph_mean), sphericity standard deviation (sph_std), median particle size ($D_{50}$), and texture. Values indicate linear correlation strength r between descriptor pairs, with color intensity denoting correlation magnitude. (b) Pairwise scatter plots with fitted regression lines and Pearson correlation coefficients r labeled.

To assess the mutual independence and relevance of selected morphology descriptors, pairwise Pearson correlation coefficients were calculated between sphericity, sphericity standard deviation, $D_{50}$, and texture. As shown in the correlation matrix (Figure S4a), weak correlations across most pairs ($|r| < 0.5$) suggest low redundancy between descriptors, supporting their combined use in statistical and ML models for precursor morphology characterization. The correlations between texture and the other descriptors remained low ($|r| \leq 0.26$), confirming that the wavelet-derived texture descriptor captures distinct aspects of microstructure not reflected in $D_{50}$ or sphericity. This statistical independence ensures that each metric contributes complementary information for downstream morphology optimization. Scatter plots with regression fits in Figure S4b further visualize these trends and confirm that no single descriptor dominates the morphological landscape. The low redundancy among descriptors validates their concurrent use in data-driven modeling frameworks.

# 5. Forward Prediction of Precursor Morphology under Diverse Synthesis Conditions

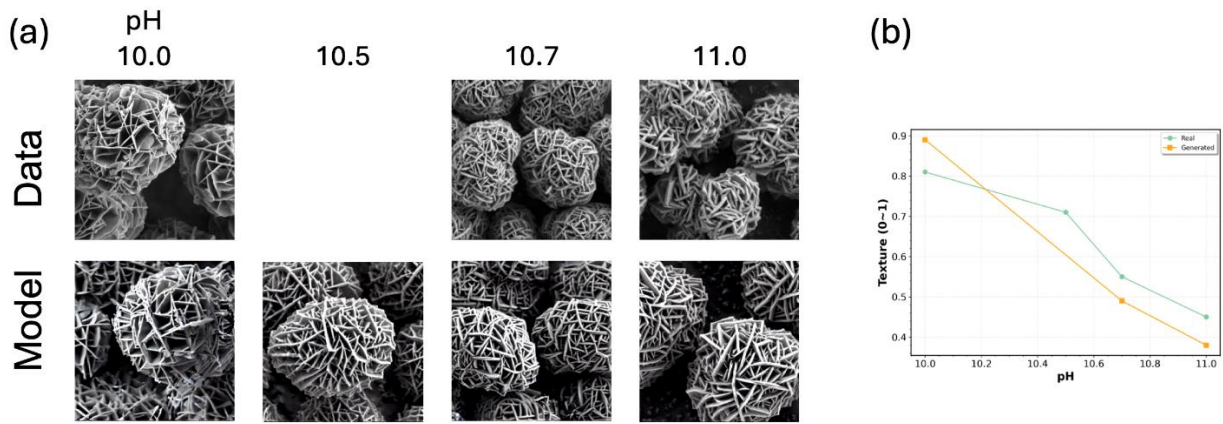

**Figure S5. pH-dependent prediction of precursor texture using the diffusion-based generative model. (a)** Comparison between experimental SEM images (top row) and model-generated images (bottom row) across four pH values. Images at pH 10.0, 10.7, and 11.0 correspond to experimental conditions used during model training. The image at pH 10.5 represents an interpolation point not seen during training, demonstrating the model's generalization capability. The generated morphologies closely resemble the experimentally observed texture evolution, including the progressive smoothing and thickening of primary particles as pH increases. **(b)** Quantitative comparison of texture values as a function of pH. The model successfully captures the monotonic decrease in texture with increasing pH, and interpolated prediction at pH 10.5 remains consistent with the experimental trend, validating the model's robustness

Figure S5 illustrates the prediction capability of our diffusion-based generative framework under variable pH conditions, capturing not only static morphological features but also the dynamic evolution of primary particle structure over time. As demonstrated in Figure 3, precursor particles exhibit significant changes in thickness, stacking, and texture over reaction time. The model successfully internalizes these temporal dynamics alongside pH trends, enabling realistic generation of intermediate morphologies that reflect both kinetic and chemical influences.

The generated images are visually and quantitatively consistent with experimentally observed textural changes even at the interpolated pH 10.5 condition. As pH increases, enhanced deprotonation of $TM(OH)_2$ structures results in more negatively charged surfaces, which in turn strengthen electrostatic attraction with positively charged metal-ammonia complexes such as $[TM(NH_3)_n]^{2+}$. This preferential interaction promotes directional growth along the [001] axis, leading to thicker, more compact primary structures. These morphological transitions are faithfully reproduced by the model.

Quantitative texture analysis confirms the model's predictive accuracy: a monotonic decline in high-frequency wavelet components aligns with experimentally observed increases in primary particle thickness and reduced porosity. This demonstrates that the model does not merely memorize training samples but learns physically meaningful correlations between synthesis parameters and structural outcomes. The ability to generate realistic morphologies under unseen synthesis conditions while preserving known physical trends validates the model's generalization capability and underscores its utility for inverse design targeting user-specified microstructural objectives.

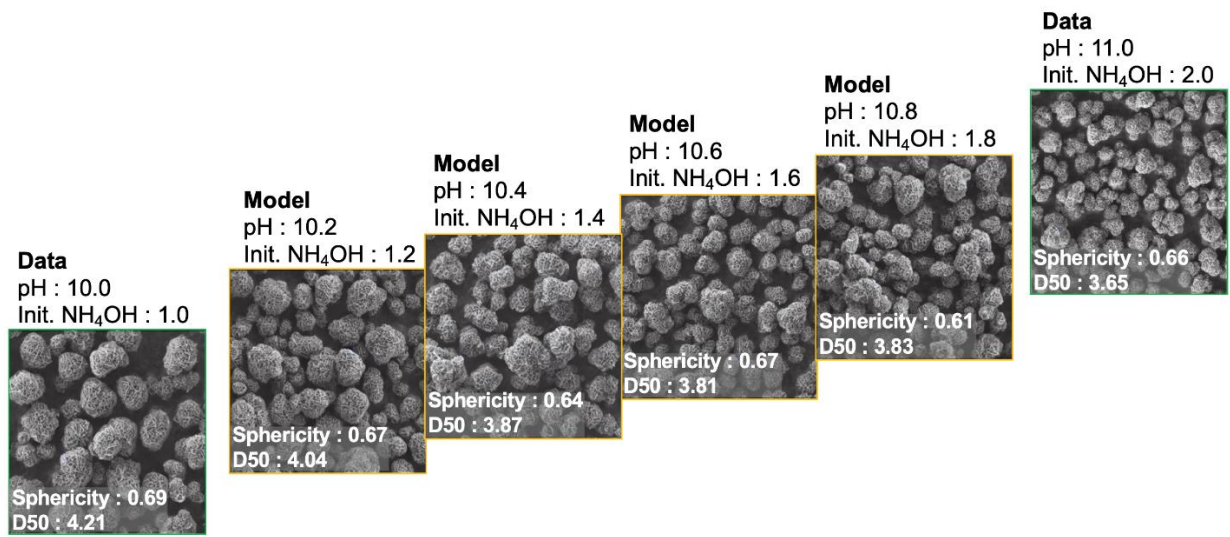

**Figure S6. Forward prediction of precursor morphology under coupled variation of pH and initial NH$_4$OH concentration.** Comparison of experimental (left and right ends) and model-generated (interpolated) SEM-like images for precursor particles at various synthesis conditions. pH was varied from 10.0 to 11.0 and initial NH$_4$OH concentration from 1.0 to 2.0 in 0.2 increments. The generative model was trained only on the end-point experimental data (pH 10.0 and 11.0) and used to predict intermediate conditions. Each image is annotated with sphericity and D$_{50}$ (median particle size) values extracted via the morphology quantification framework. The predicted morphologies generally show increased sphericity and reduced particle size with increasing pH

and $NH_4OH$ concentration, consistent with physical trends associated with surface charge modulation, enhanced complexation, and suppressed aggregation. However, small deviations from monotonic trends such as local fluctuations in sphericity and $D_{50}$ appear in the interpolated predictions, reflecting the limitations of generative interpolation across multivariate compositional spaces. Despite these minor discrepancies, the model successfully captures key structure-controlling mechanisms and demonstrates strong potential for forward design across coupled synthesis parameters.

Figure S6 extends the forward prediction capability of our diffusion model into a two dimensional synthesis space, where both pH and initial $NH_4OH$ concentration are varied simultaneously. This experiment builds upon the temporal evolution of primary particle morphology discussed in Figure 3 and the pH-dependent predictions shown in Figure S5. While Figure 3 focused on the time-dependent coarsening behavior of primary particles under fixed chemical conditions, and Figure S5 demonstrated the model's ability to interpolate along a single pH, Figure S6 evaluates the model's performance when jointly interpolating along both compositional parameters. As expected from physical chemistry principles and literature, increasing pH leads to enhanced deprotonation of hydroxide surfaces, resulting in greater negative surface charge. This promotes electrostatic repulsion between primary particles, suppressing aggregation and contributing to the formation of smaller, more isolated secondary particles. Simultaneously, high pH raises supersaturation, leading to rapid nucleation and finer particles.[19, 20]

In parallel, increasing $NH_4OH$ concentration strengthens metal-ligand complexation, which slows down precipitation kinetics and supports homogeneous nucleation. This results in more spherical and densely packed secondary morphologies. In contrast, insufficient $NH_4OH$ leads to uncontrolled nucleation and irregular morphologies due to the presence of uncomplexed $TM^{2+}$ and disrupted Ostwald ripening.[4, 21]

The generated images qualitatively follow these trends. As both pH and $NH_4OH$ increase, secondary particles become more spherical, compact, and size-controlled. However, since the intermediate data points were generated through interpolation, not experiment, some minor deviations from expected monotonic trends are observed. For example, the sphericity shows non-monotonic variation (e.g., slightly lower at pH 10.4 than at 10.2), and $D_{50}$ values at certain conditions do not decrease consistently. These fluctuations are not artifacts of experimental noise but rather reflect limitations in the model's interpolation fidelity across coupled synthesis parameters.

Such deviations highlight the complexity of multivariate synthesis-structure relationships, where nonlinear interactions such as local saturation effects, kinetic lag, or anisotropic growth modes may not be fully captured by training data sparsely distributed in the parameter space. Nonetheless, the model's ability to maintain physically reasonable trends and generate high quality SEM-like morphologies across unseen conditions validates its utility for guiding experimental design and informs further refinement of generative control.

**Supplementary Reference**